\documentstyle[aps]{revtex}

\begin{document}

\twocolumn[\hsize\textwidth\columnwidth\hsize\csname @twocolumnfalse\endcsname

\title{
The effects of varying the strengths of tensor and spin-orbit
interactions on M1 and E2 rates in $^{12}$C: Comparison of results in
$\Delta N = 0$ and $\Delta N = 0 + 2\hbar\omega$ spaces}

\author{M.S. Fayache,$^{1,2}$, Y.Y. Sharon$^{2}$\cite{byline} and
L. Zamick,$^{2}$}
\address{(1) D\'{e}partement de Physique, Facult\'{e} des Sciences de
Tunis, Tunis 1060, Tunisia\\
\noindent (2) Department of Physics, Rutgers University, New
Brunswick, NJ 08903 USA}

\maketitle

\begin{abstract}
The energies and transition rates to $J=1^{+}~T=1$ and $J=2^{+}~T=0,1$
states in $^{12}$C with matrix elements fitted to realistic $G$ matrix
elements obtained in non-relativistic approaches are studied.
Then the effects of varying the strengths of the two-body tensor and
spin-orbit interactions are also considered.  The calculations are done in
both a small space (0p) and a large space (0p + $2\hbar\omega$).  In
the small space the B(M1) from ground to the $J=1^{+}~T=1$ is
enhanced and gets closer to experiment if the strength of the
spin-orbit interaction is increased and/or if that of the tensor
interaction is made weaker.  In a large space the spin B(M1) gets reduced by
almost a factor of two.  A `self-weakening' mechanism for the tensor
interaction which succeeded in explaining anomalies in other nuclei
does not seem to work for this case.
\end{abstract}

\vspace{0.3in}
]


\section{Experiment and Previous Studies}

Of necessity, shell model calculations must be carried in limited
shell model spaces.  It is therefore of interest to try to understand
the impact of such restriction on the results that are obtained.  In
this paper we study the effects of enlarging the shell model space on
M1 and E2 excitation rate and we will also study the effects of
varying strengths of the spin-orbit and tensor interactions.

There are two well studied $J=1^{+}$ states in
$^{12}$C~\cite{Ajzenberg-Selove-F:1990npa}.
These are the $T=0$ states at 12.71~MeV and the $T=1$ state at
15.11~MeV.  The B(M1) rates to these states from ground are
respectively 0.040(3)~$\mu_{N}^{2}$ and 2.63(8)~$\mu_{N}^{2}$.
The dominance of the isovector over the isoscalar B(M1) an easily be
understood by taking the ratio of the squared isovector and isoscalar
spin $g$ factors
\begin{displaymath}
\left( {g_{s\pi} - g_{s\nu} \over g_{s\pi} + g_{s\nu}} \right)^{2}
  = \left( {9.412 \over 1.760} \right)^{2} \approx 81
\end{displaymath}
Nuclear structure effect make the ratio even larger, $\approx
336$~\cite{fnrsz}.  Because the transition to the dominantly
$J=1_{1}^{+}~T=0$ state is so weak it is strongly affected by isospin
impurities, due mainly to the Coulomb
interaction~\cite{Adelberger-EG:1976plb,Flanz-JB:1979prl,Graf-HD}.  This has
previously been considered by Sato and Zamick~\cite{Sato-H:1977plb},
where it was 
shown that the surprisingly large Coulomb matrix element needed 
by Adelberger {\it et al.}~\cite{Adelberger-EG:1976plb} to
induce the isospin mixing could be explained by taking into
account nuclear binding energy effects.  In this work, however, we
will focus of the $J=1^{+}~T=1$ states and so we use the experimental
value of 2.63(8)~$\mu_{N}^{2}$ as a point of comparison for our
calculations.

There is some interest in correlating B(M1) transitions both
orbital~\cite{Ziegler-W:1990prl,Heyde-K:1991prc,zz-prc} and spin~\cite{zza,abbz,azzb}, with
B(E2)'s i.e. with quadrupole collectivity.  We note 
that there is good evidence that $^{12}$C is strongly deformed ($\beta
\approx -0.6$).  The magnitude of $\beta$ comes from the experimental
B(E2) $0_{1} \rightarrow 2_{1}$ which equals $\sim 40$ $e^{2}$
fm$^{4}$~\cite{Raman-S:1987ndt}.

Using phenomenological interactions Cohen and
Kurath~\cite{Cohen-S:1965np} had 
no difficulty in fitting this data.  They obtained the states at
12.43~MeV and 15.23~MeV with B(M1) values of 0.0148~$\mu_{N}^{2}$
(Coulomb interaction neglected) and 2.510~$\mu_{N}^{2}$ respectively.
However, calculations by Zheng and Zamick~\cite{zz-annp}, using $G$ matrices
calculated by non-relativistic methods from realistic interactions, gave
much too small a B(M1) to the $J=1^{+}~T=1$ state at 15.11 MeV.  It
was noted that increasing the spin-orbit strength and/or weakening the
tensor interaction strength could cause the B(M1) to be increased.  A
justification for increasing the spin-orbit strength comes from the
Dirac phenomenology~\cite{Serot-BD:1986anp}.  Indeed, Zheng, Zamick
and M{\"u}ther~\cite{zzm} 
showed that with $G$ matrices derived from the relativistic Bonn A
interaction~\cite{Machleicht-R:1989anp}, the value of B(M1) $0^{+}~T=0
\rightarrow 1^{+}~T=1$ 
increased with decreasing Dirac mass.  For example, with
$m^{*}$(DIRAC) = 938.9, 729.1 and 630.0 MeV/$c^{2}$ respectively, the
values of B(M1) were 0.69, 1.13 and 1.57~$\mu_{N}^{2}$ respectively.
Zheng and Zamick had introduced the $(x,y)$ interaction~\cite{zz-annp}
\begin{equation}
V = V_{c} + x V_{{\rm s.o.}} + y V_{{\rm t}}
  \label{eq:interaction}
\end{equation}
where s.o. = spin orbit, t = tensor and c = everything else,
especially a spin-dependent central interaction.  For $x=1$, $y=1$
this interaction gave a good fit to the non-relativistic Bonn A matrix
elements.  On could simulate the effects of the Dirac phenomenology
very well by making $x$ larger than one.

One can also get a larger B(M1) by making $y$ smaller.  The
justification for this is less clear.  There are the universal
scaling ideas of G.~E. Brown and M.~Rho~\cite{Brown-GE:1991prl} where
inside the nucleus the 
masses of all mesons except for the pion decrease relative to their
free space values.  
In this model not only does the spin-orbit interaction become stronger
(as in Ref.~\cite{Serot-BD:1986anp}), but also the tensor interaction
becomes weaker.
Another possibility is the that the effect could
be due to higher shell admixtures, without having to invoke
non-nuclear degrees of freedom.  Concerning the latter, Fayache, Zheng
and Zamick~\cite{zzf} noted that higher shell admixtures, i.e. $2 \hbar
\omega$ and higher excitations, could for many phenomena make the
tensor interaction look weaker in the valence space.  They called this
the `self weakening mechanism'.  They applied this to the problems of
the quadrupole moment of the ground state of $^{6}$Li, the
near-vanishing of the Gamow-Teller matrix element for the $A=14$
system ($^{14}$O [$J=0~T=1$] $\rightarrow$ $^{14}$N [$J=1~T=0$]) and the
energy splitting of the $J=0_{1}^{-}~T=1$ and $J=0_{1}^{-}~T=0$ states
in $^{16}$O.  In all three cases the experimental result could be
obtained either by weakening the tensor interaction in a valence space
calculation i.e. making $y$ less than one, or by keeping the full
strength $y=1$ and performing shell model calculations in which $2
\hbar \omega$ and sometimes higher admixtures were included.

However, in a perturbation theory approach by Zheng, Zamick and
M{\"u}ther~\cite{zzm} to the problem of B(M1) $^{12}$C [$J=0~T=0
\rightarrow J=1^{+}~T=1$] the self-weakening mechanism did not seem to
work.  The authors renormalized the two body matrix elements in the 0p
shell and then used these in a 0p shell matrix diagonalization.  If
only Bertsch-Kuo-Brown bubble~\cite{Bertsch-GF:1965np,Kuo-TTS:1966np}
was exchanged between two nucleons, then 
indeed the B(M1) got bigger.  However, this was offset by the two
other diagrams -- hole-hole and
particle-particle~\cite{Bertsch-GF:1965np,Kuo-TTS:1966np}.  The latter
reinforce each other and cause B(M1) to become small again   The
specific number for the Bonn A interaction were 0.639~$\mu_{N}^{2}$
for the bare interaction, 1.177~$\mu_{N}^{2}$ when the bubble diagram
was included, 0.832~$\mu_{N}^{2}$ when the bubble plus hole-hole
were included and 0.514~$\mu_{N}^{2}$ when the bubble plus hole-hole
plus particle-particle diagrams were included.  We go from small to
large to smaller to smaller still.

It is not however clear if merely renormalizing the two body
interaction is sufficient for calculating a transition matrix element.
If one draws Goldstone diagrams for the M1 transition there are
several diagrams {\em not} included by merely substituting a
renormalized interaction into the shell model calculation.  Examples
of the latter include the diagrams where the M1 operator is inserted
after the first interaction but before the second interaction.
Therefore in this work we decided to do a complete valence plus $2
\hbar \omega$ shell model diagonalization for the relevant states and
the M1 transition rates in $^{12}$C.  In such an approach all the
missing diagrams are implicitly generated.

Of particular interest is whether the large space calculation will or
will not serve to justify the adjustment of the parameters $x$ and
$y$ in the small space.  For example, will and $x=1$, $y=1$ calculation
in the large space yield results similar to those of an $x=1$, $y=0.5$
calculation in the small space (self-weakening of the tensor
interaction)?  Such a comparison makes sense only if we look at the
low energy sector of the large space results.

Our calculations for the B(M1)'s are of additional interest because
of the close relationship of the spin B(M1)$_{\sigma}$'s to the beta
decay Gamow-Teller matrix elements.  Thus our studies of the effects
of enlarging the shell-model space have a broader applicability.

\section{Comparison of Results of B(M1) 
$J=0^{+}~T=0 \rightarrow J=1_{1}^{+}~T=1$ in $^{12}$C in the small space
($\Delta N = 0$) and the large space ($\Delta N = 2$)}

\subsection{The $x=1$, $y=1$ results}

In the small space (see table~\ref{tab:BM1}) with the $x=1$, $y=1$
interaction (which corresponds roughly to $G$ matrix elements of a
realistic interaction obtained in a non-relativistic approach) we obtain
for B(M1))$\uparrow$ to the lowest $J=1^{+}~T=1$ state a value of
0.89~$\mu_{N}^{2}$,
much smaller than the experimental value of 2.63(8)~$\mu_{N}^{2}$.  Our
calculation further indicates that this is dominantly a spin-mode
excitation, since B(M1)$_{l}$ = 0.03~$\mu_{N}^{2}$ and 
B(M1)$_{\sigma}$ = 0.58~$\mu_{N}^{2}$.  

Does the discrepancy between theory and experiment go away when we go
to the large space?  Quite the contrary, the B(M1) gets smaller, mainly
because B(M1)$_{\sigma}$ decreases.  Note the B(M1)$_{\sigma}$ is very
closely related to B(GT) the Gamow-Teller transition.  They will have
{\em identical quenching} due to configuration mixing.   From
table~\ref{tab:BM1} 
we see that relative to $\Delta N = 0$, the quenching factor for
B(M1)$_{\sigma}$ in a $\Delta N = 0 + 2 \hbar \omega$ calculation is
0.31/0.58 = 0.53 (or for the operator it is 0.73).

This is in accord with the general consensus of what the quenching of
Gamow-Teller operators should be.  We have here verified this result,
not in perturbation theory, but rather in an explicit matrix
diagonalization.  It appears that matrix diagonalization results are
close to those of perturbation theory.

A point that has been made before~\cite{zz-annp} is worth repeating.
In open shell 
nuclei, one can get B(M1)$_{\sigma}$'s and B(GT)'s to be too small,
despite the fact that we believe there should be an overall quenching.
This is what is happening here when we use non-relativistic  $G$
matrixes derived from realistic interactions.  As another example, in the
SU(4) limit for $N=Z$ nuclei B(GT) will vanish.  This limit can be
reached by using a spin-independent two-body interaction and turning
off the one-body spin-orbit interaction, e.g. in the Elliott model
where one has only a $\rm Q \cdot Q$
interaction~\cite{Elliott-JP:1958prsla}.  Thus in a given 
calculation in an open shell nucleus one cannot readily deduce
where one needs quenching or enhancement of the spin, magnetic
and/or Gamow-Teller operator, unless one has complete confidence in
the two-body interaction and single particle energies that are being
used.


In table~\ref{tab:BM1-summed}(a) we present the summed M1 strengths in
the small space; 
in table~\ref{tab:BM1-summed}(b) we have the summed strengths to the
first 10 states in 
a large space calculation ($\Delta N = 0 + 2\hbar\omega$) and in
table~\ref{tab:BM1-summed}(c) the sum to 500 states.

Whereas for the lowest state in the small space calculation
(table~\ref{tab:BM1}) 
for the $x=1$, $y=1$ interaction the values of B(M1), 
B(M1)$_{l}$ and B(M1)$_{\sigma}$ were respectively 0.89, 0.03 and
0.58~$\mu_{N}^{2}$, the summed small space values are 1.42, 0.60 and
0.91~$\mu_{N}^{2}$.  In particular there is some isovector orbital
(scissors mode) strength in higher $J=1^{+}~T=1$ states.

Focusing on B(M1)$_{\sigma}$ we see that the small space sum of
0.91~$\mu_{N}^{2}$ becomes 0.70~$\mu_{N}^{2}$ when we consider the
first 10 states in a large space calculation. There is quenching of
the low lying strength.  
We will compare the small and large space total strengths in a later section.

\subsection{Varying $x$ and $y$}

In tables~\ref{tab:BM1} and~\ref{tab:BM1-summed} we study also the
variation of the B(M1) rates 
with $x$ and $y$, the strengths of the two-body spin-orbit and tensor
interactions respectively.  We consider four sets of $(x,y)$ -- (1,1),
(1.5,1), (1,0.5) and (1.5,0.5).  Increasing the spin-orbit strength
from $x=1$ to $x=1.5$ simulates to a large extent the use of Dirac
phenomenology with a Dirac effective mass of 1/1.5 = 0.67.  If indeed
the source of the increase in the spin-orbit interaction comes from
the Dirac phenomenology, then one is justified in using this enhanced
value of $x$ in both the small space ($\Delta N = 0$) and the large
space ($\Delta N = 0+2 \hbar\omega$).

For the tensor interaction the choice of the parameter $y$ in the
large space calculation depends which of the scenarios discussed in
the previous section for weakening the tensor interaction are
correct.  If the universal scaling ideas of G.~E. Brown and
M.~Rho~\cite{Brown-GE:1991prl} are correct 
then one should use a weaker tensor interaction, e.g. $y=0.5$ in both
the small space and large space calculation.  However in the
`self-weakening mechanism' which explained several phenomena for
nuclei with mass numbers $A$ = 6, 14 and 16, the higher shell
admixtures were responsible for 
making the tensor interaction appear weaker in the valence space.  In
that case if one uses $y=0.5$ in the small spaces one should use
$y=1$ in the large space and hope that the results for the low energy
properties are about the same in the two calculations.

First, focusing on the small space calculation of table~\ref{tab:BM1}
we see that 
we get a marked improved in the B(M1) rate $J=0^{+}~T=0 \rightarrow
J=1^{+}~T=1$ when we increase the spin-orbit interaction and/or
decrease the tensor interaction.  From $(x,y)=(1,1)$ to (1.5,0.5) the
value of B(M1) increase from 0.89 to 2.54 and the value of B(M1)$_{\sigma}$
increase from 0.58 to 2.57.  We are getting close to the experimental
value of 2.68~$\mu_{N}^{2}$.

In the large space the values of B(M1) to the first $1^{+}~T=1$ state
are however in all cases 
smaller than in the small space.  The `self-weakening mechanism' for
the tensor interaction does not appear to work here.  We should
compare the $(x,y)=(1.5,1)$ large space B(M1) result of
1.29~$\mu_{N}^{2}$ with the $(x,y)=(1.5,0.5)$ small space result of
2.54~$\mu_{N}^{2}$.  For $(x,y)=(1.5,1)$ the gains from the small
space to the large space does not increase B(M1) -- rather it {\rm
decreases} it from 1.89~$\mu_{N}^{2}$ to 1.29~$\mu_{N}^{2}$.
This is mainly due to a decrease of B(M1)$_{\sigma}$ from
1.71~$\mu_{N}^{2}$ to 0.98~$\mu_{N}^{2}$ (B(M1)$_{l}$ changes from
0.04 to 0.02).
Why the `self-weakening mechanism' appears to work in some nuclei but
not in others is not clear and bears further investigation.

It should be added that results we obtain here via matrix
diagonalization are not so different from those by Zheng, Zamick and
M{\"u}ther~\cite{zzm} using a renormalized interaction in perturbation
theory.  Evidently, the missing diagrams in which the M1 operator acts
{\em between} the two interactions are not so important.


We see from table~\ref{tab:BM1-summed}, which deals with {\em summed}
B(M1) strengths, that in 
the small space increasing the spin-orbit 
strengths from $x=1$ to $x=1.5$ results in very significant changes in
the summed isovector $M1$ excitation strengths.  More specifically,
with this change in $x$, B(M1) is increased by over 60\%,
B(M1)$_{\sigma}$ is increased by 127\% and B(M1)$_{l}$ is decreased by
16\%.  The effect in the small space of decreasing the strength of the
tensor force from $y=1$ to $y=0.5$ is similar but smaller; B(M1) is
increased by 11\%, B(M1)$_{\sigma}$ is increased by 24\% and
B(M1)$_{l}$ is decreased by 1.6\%.  The combined effect of increasing
the spin-orbit and decreasing the tensor strengths is more than the
sum of the individual effects.  Here B(M1) is doubles,
B(M1)$_{\sigma}$ is more than tripled which B(M1)$_{l}$ decreases by
23\%.

From table~\ref{tab:BM1-summed} we also see that in the large space
the effect on the 
isovector B(M1) strength of increasing strength $x$ of the
spin-orbit force is qualitatively similar to the effect of doing so in
the small space, but is smaller in scale by a factor of about 0.6.
When $x$ is increased fro 1 to 1.5 (both with $y=1$), the B(M1) is
increased by 36\%, B(M1)$_{\sigma}$ is increased by 80\% and
B(M1)$_{l}$ is reduced by 10\%.  Very interestingly, in the large
space, changing only the strength $y$ of the tensor force has
essentially no effect on the summed isovector M1 strength.  Indeed, when
$x=1$ and $y$ is changed for 1 to 0.5, then the B(M1),
B(M1)$_{\sigma}$ and B(M1)$_{l}$ all change by less than 3\%.  This
result requires further study.  On the other hand, in the large space,
the effect on the isovector B(M1)'s of decreasing the tensor
strength $y$ (from 1 to 0.5) after the spin-orbit strength has already
been increased for $x=1$ to $x=1.5$ is much more substantial.  It
results in a further increase of 17\% in B(M1) and close to 30\% in
B(M1)$_{\sigma}$ and a further decrease of 8\% in B(M1)$_{l}$.

The middle part of table~\ref{tab:BM1-summed} shows for any
given $(x,y)$ 
combination what happens if we limit ourselves to the first ten states
in the large space.  Then we typically underestimate the isovector
B(M1), B(M1)$_{\sigma}$, B(M1)$_{l}$ values by 10\% to 30\% in
comparison with the results obtained by summing over the lowest 500
states in the large space.  In the small space there are only eight
states with $J=1^{+}~T=1$.  For any one $(x,y)$ combination the
corresponding values of B(M1) and B(M1)$_{\sigma}$ are typically
20-40\% larger than for the ten lowest states in the large space, but
the B(M1)$_{l}$'s hardly change (by less than 7\% in all cases.

\subsection{Comparison of the total isovector M1 strength in the small
and large spaces}

By including $2\hbar\omega$ excitations in our shell model
diagonalization we are able to address the question of whether or not
the isovector M1 strength which disappears from the low lying sector
(and in particular from the lowest $1^{+}~T=1$ state at 15.11 MeV)
reappears in the high energy sector, i.e. in the $2\hbar\omega$
region.  To answer this we merely have to look at the results in
table~\ref{tab:BM1-summed}.

For the case $x=1, y=1$ the total strength in the 0p$+2\hbar\omega$
space is larger than in the small space.  For the other three cases
$(x,y)=(1.5,1),(1,0.5)$ and $(1.5,0.5)$ the opposite is true -- the
total isovector magnetic dipole strength is smaller when
$2\hbar\omega$ excitations are included.

For  $x=1, y=1$ the summed B(M1) strength in the small space is
1.42~$\mu_{N}^{2}$ whereas in the large space it is
1.61~$\mu_{N}^{2}$.  In this case, where there is quenching in the
low-lying sector this is more than compensated for by the strengths in
the $2\hbar\omega$ region.

For the other three cases, the reverse is true.  For example in the
case $x=1.5, y=0.5$ the summed strength in the 0p space is
2.58~$\mu_{N}^{2}$, whereas in the 0p$+2\hbar\omega$ space it is
2.56~$\mu_{N}^{2}$.  If we look at B(M1)$_{\sigma}$ the drop in the
sum is larger, from 2.82~$\mu_{N}^{2}$ to 2.28~$\mu_{N}^{2}$, a
decrease of 19\%.  The orbital summed strength B(M1)$_{l}$ on the
other hand increases for this case from 0.46~$\mu_{N}^{2}$ to
0.57~$\mu_{N}^{2}$.  Indeed the orbital summed strength increases for
all four $(x,y)$ combinations considered here.

From the above we see that the M1 strength is redistributed but not
conserved.  There is definately a quenching in the low energy sector
but in three out of four cases considered here the missing strength
does not reappear in toto in the high energy sector.

\section{Comparison of Results of B(E2) 
in $^{12}$C in the small space
($\Delta N = 0$) and the large space ($\Delta N = 2$)}

As a counterpoint to the high lying M1 transitions we wish to now
study what happens to low lying states, e.g. the $2_{1}^{+}$ state,
when we increase the model space and when $x$ and $y$ are varied.
In the small space (0p) we use effective charges $e_{p}=1.5$,
$e_{n}=0.5$.  In the large space (0p+$2\hbar\omega$) we use bare
charges $e_{p}=1$, $e_{n}=0$.  The $\Delta N = 2\hbar\omega$
admixtures should in principle at least justify the use of effective
charges in the small space.
The results of the B(E2)'s to individual states are presented in
table~\ref{tab:BE2} and the summed strengths in table~\ref{tab:BE2-summed}.

\subsection{The $x=1, y=1$ results}

\subsubsection{Isoscalar B(E2) --
$J=0^{+}~T=0 \rightarrow J=2^{+}~T=0$}

In the small space there are eight $J=2^{+}~T=0$ states.  In that
space (with the effective charges $e_{p}=1.5, e_{n}=0.5$), 95\% (80.6
$e^{2}$ fm$^{4}$) of the total calculated B(E2) strength of 82.7 
$e^{2}$ fm$^{4}$ is concentrated in the transition to the lowest
$J=2^{+}~T=0$ state at 3.80 MeV.  In the larger space, the lowest
$J=2^{+}~T=0$ state now at 4.79 MeV is still dominant with a B(E2)
of 23.2 $e^{2}$ fm$^{4}$ and accounts for 55\% of the total strength
of 42.5 $e^{2}$ fm$^{4}$.  In addition, another relatively strong
$J=2^{+}~T=0$ state emerges among the ten lowest state, its energy is
4.40 MeV and its B(E2) is 6.12 $e^{2}$ fm$^{4}$.  The B(E2)$_{{\rm
isoscalar}\uparrow}$ sum, over the first 500 states in the large
space, is slightly less than one half of the total B(E2)$_{\uparrow}$
sum over all the states in the small space calculation.  The sum over
the first ten states in the large space (up to about 45 MeV of
excitation) adds up to 31.9 $e^{2}$ fm$^{4}$ or about 3/4 of the sum
over the first 500 states.

The dominance of one single state suggest collectivity.  In $N=Z$
nuclei the isoscalar B(E2) is proportional to $(e_{p}+e_{n})^{2}$.
Usually the values of the effective charges $e_{p}, e_{n}$ are
introduced in $\Delta N = 0$ calculations to reproduce the effects of
neglecting higher shells.  The comparison of the large space and small
space results suggests that $e_{p}+e_{n}$ should be more like 1.4
instead of $1.5+0.5=2$.  On this point it was noted by Abbas and
Zamick~\cite{Abbas-A:1980prc} that the isoscalar effective charges is
less than two for light nuclei and only reaches the value of two
asymptotically for large $A$.

\subsubsection{Isovector B(E2) --
$J=0^{+}~T=0 \rightarrow J=2^{+}~T=1$}

In general this isovector B(E2) sum is smaller than the
corresponding isoscalar one.  In the small space, with a total of
seven $J=2^{+}~T=1$ states the isovector B(E2)$_{\uparrow}$ sum is
dominated by the second $J=2^{+}~T=1$ state (here at 15.78 MeV), which
contributes 2.61 $e^{2}$ fm$^{4}$ out of a total value of 3.39 $e^{2}$
fm$^{4}$ for the B(E2) sum.  In the large space calculation the
largest contribution to the isovector B(E2) sum from among the few
lowest-lying state again comes from the second excited $J=2^{+}~T=1$
state now at 20.08 MeV above the ground state with a B(E2) of 1.56
$e^{2}$ fm$^{4}$.  The sum over the B(E2)$_{\uparrow}$'s to the first
ten $J=2~T=1$ states (which corresponds to an excitation energy of up
to 53 MeV) is only 2.91 $e^{2}$ fm$^{4}$, but when the 500 lowest
states are taken into account the B(E2) sum increases about
seven-fold to 21.28 $e^{2}$ fm$^{4}$.  It would thus seem that there
are substantial isovector B(E2) contribution from very high lying
states.  The isovector B(E2)'s for $N=Z$ nuclei are proportional to
$(e_{p}-e_{n})^{2}$; reductions in both $e_{p}$ and $e_{n}$ will not
necessarily reduce this quantity, unless such reduction are carried
out in such a way that the difference $e_{p}-e_{n}$ is also reduced.
There is evidence that indeed $e_{p}-e_{n}$ should be less than unity,
from theoretical
calculations~\cite{Zamick-L:1969prev,Bohr-A:1975-nucl-struct} and from
phenomenological
fits~\cite{Wilderthal-BH:1984ppnp,Brown-BA:1988arnps}. 
A popular value is 0.7. 

\subsection{Varying $x$ and $y$}

\subsubsection{Isoscalar B(E2) --
$J=0^{+}~T=0 \rightarrow J=2^{+}~T=0$}

We see from table~\ref{tab:BE2} the dominant B(E2)'s, the isoscalar ones,
depend only weakly on the particular $(x,y)$ combination that is used.
For each of the three main blocks in table~\ref{tab:BE2} (small space, large
space, the lowest states in the large space) the variation of the
isoscalar B(E2) among the different $(x,y)$ is less than 9\%.
Typically , with $x=1$ decreasing the tensor force from $y=1$ to
$y=0.5$ has a negligible decreasing effect (less than 1\%). Increasing
the spin-orbit from $x=1$ to $x=1.5$ both with $y=1$ decreasing the
isoscalar B(E2) by 3-7\%, and simultaneously decreasing the tensor
and increasing the spin-orbit decreases the isoscalar B(E2) by a
total of 5 to 9\%.

The isoscalar B(E2) does depend critically, however on the space in
which the calculation is carried out and other effective charges used.
The approximate B(E2) values in $e^{2}$ fm$^{4}$ are 90- for the
small space (with $e_{p}=1.5, e_{n}=0.5$), 30 for the first ten
$2^{+}$ states in the large spaces and 40 for the large space.  We
thus see that the isoscalar B(E2) sum decreases as we calculate it in
larger spaces (with the actual physical charges $e_{p}=1, e_{n}=0$).

Several conclusions can be drawn from the results which were outlined
in the previous paragraph.  The isoscalar B(E2)'s for $N=Z$ nuclei
are proportional to $(e_{p}+e_{n})^{2}$.  We use effective charges
$e_{p}=1.5$ and $e_{n}=0.5$ in the small space to compensate for the
non-inclusion of higher shells.  These charges lead to B(E2)'s that
are twice as large in the $\Delta N = 0$ calculation as in the
$\Delta N = 2$ calculation.  If we use the large ($\Delta N =2$) space
calculation as our standard, it would seem the effective charges
$e_{p}+e_{n} = 2$ used in the small space are too large and that this
sum should be smaller by a factor of about 1.4 (i.e. $e_{p}+e_{n}
\approx 1.4$) at least for $^{12}$C and possibly more general in the
$0p$ shell.

The use of effective charges that are too large in the small space
($\Delta N = 0$) calculations leads to exaggerated B(E2)'s and hence
to exaggerated deformations.  Such a calculation makes $^{12}$C appear
more deformed than it really is.

In the small space at least 96\% of the dominant isoscalar B(E2)
sum is due to one state, the lowest $J=2^{+}~T=0$ state which is
calculated to have an excitation energy of about 3.8-4.1 MeV (for the
different $(x,y)$ combinations).

In the large space that state moves to 4.6-4.9 MeV (for the different 
$(x,y)$ combinations) and its isoscalar B(E2) value increases but
only 10-12\%; however, now for all the $(x,y)$ combinations there is
another strong state (with strength 20-30\% that of the first excited
state), which lies high at 42-44 MeV, and is the eight or ninth
excited $J=2^{+}~T=0$ state.  Together, these two strong states supply
about 70\% of isoscalar B(E2)$_{\uparrow}$ strength summed over the
first 500 $J=2^{+}~T=0$ states in the large space.

\subsubsection{Isovector B(E2) --
$J=0^{+}~T=0 \rightarrow J=2^{+}~T=1$}

Here there is a marked difference in the isovector B(E2) values
between small and large spaces.  The small space isovector values in
$e^{2}$ fm$^{4}$ range from about 3-5 (depending on the $(x,y)$
values) while in the large space the sums cluster between 21.2 and 21.4
for all of our $(x,y)$ combinations.  In the small space there are
only seven states with $J=2^{+}~T=1$ with the third state at about 18
MeV supplying 65-80\% of the total isovector B(E2) strength.

The great increase in the summed isovector B(E2) value when one
goes to the large space is {\em not} due to the few lowest-lying
$J=2^{+}~T=1$ states.  In the lowest part of table~\ref{tab:BE2} we see that in
the ten lowest $J=2^{+}~T=1$ states in the large space we obtain an
isovector B(E2)$_{\uparrow}$ sum that is 15-20\% less than our sum in
the small space and is less than 1/5 of the total summed B(E2)
strength over the first 500 states in the large space.  it will be
interesting to see if there are one or a few very strong very high
lying $J=2^{+}~T=1$ states or whether this increase is a cumulative
effect of many states.  Another possibility is that we are getting
into the region of spurious states.

In the small space (and also for the ten low-lying states of the large
space) the summed B(E2) increasing by 5\% when the tensor force is
decreased, increases by 22-33\% when the spin-orbit is increased, and
by 30-40\% when both of the above effects take place.
Very interestingly, however, in the large space, the summed isovector
B(E2) is totally insensitive to the $(x,y)$ values, varying by less
than 1\% among all the $(x,y)$ combinations.

\section{Additional Remarks}

The main problem addressed in this work in the fact that in a
straightforward 0p calculation with realistic $G$ matrix elements
obtained in a non-relativistic calculation the computed value of
B(M1)$\uparrow$ from ground the $J=1^{+}~T=1$ state
(0.89~$\mu_{N}^{2}$) is much lower than the experimental value
2.63~$\mu_{N}^{2}$.  This was already noted in the works of Zheng {\it
et al.}~\cite{zz-annp,zzm}. It should be emphasized that in those
works, as well as in this one, the {\em single particle energies} are
calculated with the same interaction that is used between valence
nucleons. 

In the small space the B(M1) can be enhanced by making the spin-orbit
interaction stronger~\cite{zz-annp,zzm} and this can be justified by
the Dirac phenomenology~\cite{Serot-BD:1986anp}.  The situation with
the tensor 
interaction is a bit more complicated.  Since the Dirac phenomenology
in its original version~\cite{Serot-BD:1986anp} does not include
pions, the tensor 
interaction is not affected.  However, in the universal scaling
mechanism of G.~E. Brown and M.~Rho~\cite{Brown-GE:1991prl} all mesons
except the 
pion become less massive inside the nucleus as compared with free
space.  This leads to a weaker tensor interaction inside the nucleus.
This would be helpful for enhancing the value of B(M1)$\uparrow$ to
the $J=1^{+}_{1}~T=1$ state.

However, for other nuclei -- $^{6}$Li, $A=14$ and $^{16}$O -- another
mechanism for causing the tensor interaction to become weaker was
given by Zamick, Zheng and Fayache~\cite{zzf}.  This was simply to
allow $\Delta N = 2$ or higher admixtures into the valence space,
i.e. do larger shell model calculations or include the appropriate
Goldstone diagrams in perturbation theory.

Whereas the quadrupole moment of  the deuteron is positive, that of the
$J=1^{+}~T=0$ ground state of $^{6}$Li is negative ($-0.082$ $e$
fm$^{2}$).   In a (0p)$^{2}$ calculation {\em without} a tensor interaction,
$\rm Q$ will be positive.  With the realistic Nijm II $G$
matrix~\cite{Nijm2} the value of $\rm Q$ is calculation to be $-0.360$ $e$
fm$^{2}$ -- too negative, indicating the tensor interaction is too
strong.  However, when $2 \hbar \omega$ admixtures enter $\rm Q$
becomes $-0.251$ $e$ fm$^{2}$, and with up to $4\hbar\omega$
admixtures $\rm Q$ is calculated to be $-0.0085$ $e$ fm$^{2}$.  Thus
it appears that high shell admixtures cause the tensor interaction to
appear weaker in the valence space.

In a similar vein, for the $A=14$ beta decay $^{14}$C [$J=0^{+}~T=1$]
$\rightarrow$ $^{14}$N [$J=1^{+}~T=0$] the experimental Gamow-Teller
matrix element is very close to zero.  It was shown by B.~Jancovici
and I.~Talmi~\cite{Jancovici-B:1954prev} (see also
D.~R. Inglis~\cite{Inglis-DR:1953rmp})
that one needs a tensor interaction in the (0p) shell space in order 
to get a vanishing matrix element.  Again with the Nijm II
interaction, which has a tensor part, the $0\hbar\omega$ gives B(GT)
= 3.967 -- far from zero.  With $2 \hbar\omega$ admixtures this gets
reduce to 1.795.  The effective tensor interaction in the valence
space is weaker.  If further we enhance the spin-orbit interaction by
about 50\% (as per discussion in the previous section) we can get
B(GT) to be zero.

In $^{16}$O the isospin splitting of the $J=0^{+}~T=1$ and the
$J=0_{1}^{-}~T=0$ states is 1.845 MeV.  In the absence of a tensor
interaction these two states are nearly degenerate.  With Nijm II the
splitting becomes too large, 2.81 MeV.  In zero these two states are
one particle, one hole state -- admixture of the configurations
(1s$_{1/2}$ 0p$_{1/2}^{-1}$) and (0d$_{3/2}$ 0$p_{3/2}^{-1}$).  When
over and above these basic configurations we allow $2\hbar\omega$
excitations the energy splitting of $T=1$ and $T=0$ is reduced from
2.81 MeV to 1.93 MeV, much closer to experiment.  Again it appears
that higher shell admixtures cause the tensor interaction to appear weaker.

How do the higher shell admixtures affect the present problem
$^{12}$C?  Logically, if we expect that higher shell admixtures causes
the tensor interaction to be weaker, we should do the large space
calculation with $y=1$, i.e. with the `bare' tensor interaction.
However, as seen in table~\ref{tab:BM1}, the $y=1$ calculations in the
large space 
yields smaller values of B(M1) than do those in the small space.  For
example for $x=0$ the (small,large) results are 0.89~$\mu_{N}^{2}$ and
0.54~$\mu_{N}^{2}$ (recall that experiment is 2.63(8)~$\mu_{N}^{2}$).
For $x=1.5$ i.e. an enhanced spin-orbit interaction the corresponding
values are 1.89~$\mu_{N}^{2}$ and 1.29~$\mu_{N}^{2}$.

The above results should not be totally unexpected.  The isovector
spin B(M1)$_{\sigma}$ is proportional via an isospin
rotation to B(GT), the Gamow-Teller transition from ($N,Z$)=(6,6) to
(7,5) or (5,7), and for GT transitions it has been noted that a
quenching factor is generally needed.  A popular choice is 0.5 for
B(GT) or 0.7 for the matrix element M$_{\rm GT}$.  Looking at
B(M1)$_\sigma$ is table~\ref{tab:BM1} we see that for $x=1, y=1$ it gets reduced from
0.58~$\mu_{N}^{2}$ to 0.31~$\mu_{N}^{2}$ when $2\hbar\omega$
admixtures are allowed.  This is almost the factor of two that is often
quoted.

So basically we are getting a quenching rather than the enhancement
that we need.  It would appear that the only way to get an enhancement
towards the experimental value is to make the strength of the
spin-orbit interaction $x$ even larger and/or to invoke the universal
scaling argument of G.~E. Brown and M.~Rho~\cite{Brown-GE:1991prl} to
make the tensor interaction weaker.  Which, or how much of these two
mechanisms should be invoked will require further investigation.

\acknowledgements
This work was supported by a Department of Energy grant
DE-FG02-95ER-40940 and by a Stockton College summer R@PD research
grant.  We thank R. Machleicht for a useful communication.

\hsize\textwidth\columnwidth\hsize\csname @twocolumnfalse\endcsname
\mediumtext


\begin{table}
\caption{
Comparison of the results of $\Delta N = 0$ (small space) and $\Delta
N = 0 + 2\hbar\omega$ (large space) calculations of the value of B(M1)
from the ground state of $^{12}$C to the lowest (and most strongly
excited) $J=1^{+}~T=1$ state.  The results are presented for different
combinations of the spin-orbit ($x$) and tensor ($y$) strengths in the
realistic interaction of Eq.~(\ref{eq:interaction}).}
\label{tab:BM1}
\begin{tabular}{ccccccc} \hline
small space & $x$ & $y$ & $E$(MeV) & B(M1)\tablenotemark[1] 
            & B(M1)$_l$\tablenotemark[2] & B(M1)$_\sigma$\tablenotemark[3] \\
 & 1   & 1   & 13.60 & 0.89 & 0.03 & 0.58 \\
 & 1.5 & 1   & 13.08 & 1.89 & 0.04 & 1.71 \\
 & 1   & 0.5 & 13.33 & 1.12 & 0.01 & 0.90 \\
 & 1.5 & 0.5 & 13.11 & 2.54 & 0.00 & 2.57 \\ \hline
large space & $x$ & $y$ & $E$(MeV) & B(M1)\tablenotemark[1] 
            & B(M1)$_{l}$\tablenotemark[2] & B(M1)$_\sigma$\tablenotemark[3] \\
$\Delta N = 0 + 2\hbar\omega$ &
   1   & 1   & 17.28 & 0.54 & 0.06 & 0.31 \\
 & 1.5 & 1   & 16.46 & 1.29 & 0.02 & 0.98 \\
 & 1   & 0.5 & 16.84 & 0.88 & 0.026 & 0.60 \\
 & 1.5 & 0.5 & 16.28 & 1.96 & 0.026 & 1.83 \\ \hline
Experiment\tablenotemark[4] & & & 15.11 & 2.63(8) & & \\
\end{tabular}
\tablenotemark[1]{$g_{l\pi} = 1$, $g_{l\nu} = 0$, $g_{s\pi} = 5.586$,
$g_{s\nu} = -3.726$.} \\
\tablenotemark[2]{$g_{l\pi} = 0.5$, $g_{l\nu} = -0.5$, $g_{s\pi} = 0$,
$g_{s\nu} = 0$.} \\
\tablenotemark[3]{$g_{l\pi} = 0$, $g_{l\nu} = 0$, $g_{s\pi} = 0.5$,
$g_{s\nu} = -0.5$.} \\
\tablenotemark[4]{From Ref.~\cite{Graf-HD}.  Also the isoscalar
transition to the $J=1^{+}~T=0$ state at 12.71 MeV has a strength
B(M1)=0.040(3)~$\mu_{N}^{2}$.}
\end{table}

\begin{table}
\caption{Same as table I, but for the summed strengths.  In the small
space, the sum is over all eight states.}
\label{tab:BM1-summed}
\begin{tabular}{cccccc} \hline
small space\tablenotemark[1] 
  & $x$ & $y$ & B(M1) & B(M1)$_{l}$ & B(M1)$_{\sigma}$ \\
$\Delta N = 0$
 & 1   &   1 & 1.42 & 0.60 & 0.91 \\
 & 1.5 &   1 & 2.29 & 0.50 & 2.07 \\
 & 1   & 0.5 & 1.58 & 0.59 & 1.13 \\
 & 1.5 & 0.5 & 2.85 & 0.46 & 2.82 \\ \hline
large space\tablenotemark[2]
  & $x$ & $y$ & B(M1) & B(M1)$_{l}$ & B(M1)$_{\sigma}$ \\
$\Delta N = 0 + 2\hbar\omega$
 & 1   & 1   & 1.21 & 0.56 & 0.70 \\
 & 1.5 & 1   & 1.77 & 0.49 & 1.43 \\
 & 1   & 0.5 & 1.31 & 0.56 & 0.83 \\
 & 1.5 & 0.5 & 2.27 & 0.46 & 2.085 \\ \hline
large space\tablenotemark[3]
  & $x$ & $y$ & B(M1) & B(M1)$_{l}$ & B(M1)$_{\sigma}$ \\
$\Delta N = 0 + 2\hbar\omega$
 & 1 & 1 & 1.61 & 0.69 & 0.97 \\
 & 1.5 &   1 & 2.19 & 0.62 & 1.74 \\
 & 1   & 0.5 & 1.57 & 0.685 & 0.98 \\
 & 1.5 & 0.5 & 2.56 & 0.57 & 2.28 \\ \hline
\end{tabular}
\tablenotemark[1]{sum over all eight $J=1^{+}~T=1$ states in the $\Delta N = 0$
space.} \\
\tablenotemark[2]{sum over all first ten $J=1^{+}~T=1$ states in the
$\Delta N = 2$ space.} \\
\tablenotemark[3]{sum over the first 500 $J=1^{+}~T=1$ states in the
$\Delta N = 2$ space.}
\end{table}

\begin{table}
\caption{Comparison of experimental B(E2) values to the first $2^{+}$
state in $^{12}$C with shell model calculations for different
combinations of spin-orbit ($x$) and tensor ($y$) interaction
strengths.  Effective charges $e_{p}=1.5$, $e_{n}=0.5$ are used for
$\Delta N = 0$ calculations. Bare charges $e_{p}=1.0, e_{n}=0.0$ are
used for $\Delta N = 0+2\hbar\omega$ calculations.}
\label{tab:BE2}
\begin{tabular}{ccccccc} \hline
  & $x$ & $y$ & $E$ (MeV) & B(E2) ($e^{2}$ fm$^{4}$) 
              & $E$ (MeV) & B(E2) ($e^{2}$ fm$^{4}$) \\ \hline
Experiment\tablenotemark[1] 
 &     &     & 4.44 & 39.9(2.2) & 16.11 & 2.0(0.2) \\ \hline
small space
 & 1   & 1   & 3.80 & 80.6 & 15.78 & 2.6 \\
$\Delta N = 0$
 & 1.5 & 1   & 3.82 & 75.3 & 14.44 & 3.1 \\
$e_{p} = 1.5, e_{n} = 0.5$
 & 1   & 0.5 & 3.79 & 80.1 & 14.86 & 1.2 \\
 &     &     &      &      & 15.58 & 1.6 \\
 & 1.5 & 0.5 & 4.11 & 72.9 & 14.25 & 3.8 \\ \hline
large space
 & 1   & 1   & 4.79 & 23.2 & 20.08 & 1.6 \\ 
$\Delta N = 0 + 2\hbar\omega$
 & 1.5 & 1   & 4.63 & 21.7 & 18.31 & 1.1 \\
$e_{p} = 1, e_{n} = 0$
 &     &     &      &      & 19.12 & 1.2 \\
 & 1   & 0.5 & 4.69 & 23.2 & 19.49 & 1.5 \\
 & 1.5 & 0.5 & 4.87 & 21.2 & 17.94 & 2.3 \\ \hline
\end{tabular}
\tablenotemark[1]{From Ref.~\cite{Adelberger-EG:1976plb}.} \\
\tablenotemark[2]{From Ref.~\cite{Ajzenberg-Selove-F:1990npa}.}
\end{table}

\begin{table}
\caption{Comparison of the results in $\Delta N = 0$ (i.e. small) and
$\Delta N = 2$ (i.e. large) spaces of shell model calculations for
the sums of B(E2) transitions from the ground state of $^{12}$C to
$J=2^{+}~T=0$ states.  The results are presented for different
combinations of spin-orbit ($x$) and tensor ($y$) strengths in the
realistic interaction Eq.~(\ref{eq:interaction}).}
\label{tab:BE2-summed}
\begin{tabular}{ccccc} \hline
 & $x$ & $y$ & isoscalar B(E2)\tablenotemark[7] & isovector B(E2) \\ 
 &     &     & $J=0^{+}~T=0 \rightarrow J=2^{+}~T=0$
             & $J=0^{+}~T=0 \rightarrow J=2^{+}~T=1$ \\ \hline
small space 
 & 1   &   1 & 82.7\tablenotemark[1] & 3.39\tablenotemark[2] \\
$\Delta N = 0$
 & 1.5 &   1 & 78.4 & 4.44 \\
total sum
 & 1   & 0.5 & 82.0 & 3.54 \\
$e_{p}=1.5, e_{n} = 0.5$
 & 1.5 & 0.5 & 75.3 & 4.80 \\ \hline

large space
 & 1   & 1   & 31.9\tablenotemark[3] & 2.91\tablenotemark[4] \\
$\Delta N = 0 + 2 \hbar\omega$
 & 1.5 & 1   & 29.8 & 3.55 \\
$e_{p}=1, e_{n}=0$
 & 1   & 0.5 & 31.8 & 3.08 \\
 & 1.5 & 0.5 & 29.6 & 3.87 \\ \hline

large space
 & 1 & 1     & 42.5\tablenotemark[5] & 21.23\tablenotemark[6] \\
$\Delta N = 0 + 2 \hbar\omega$
 & 1.5 &   1 & 41.1 & 21.40 \\
$e_{p}=1, e_{n}=0$
 & 1   & 0.5 & 42.1 & 21.24 \\
 & 1.5 & 0.5 & 39.9 & 21.36 \\ \hline
\end{tabular}
\tablenotemark[1]{sum over all eight $J=2^{+}~T=0$ states in the
$\Delta N = 0$ space.} \\
\tablenotemark[2]{sum over all eight $J=2^{+}~T=1$ states in the
$\Delta N = 0$ space.} \\
\tablenotemark[3]{sum over all first ten $J=2^{+}~T=0$ states in
the $\Delta N = 2$ space.} \\
\tablenotemark[4]{sum over all first ten $J=2^{+}~T=1$ states in
the $\Delta N = 2$ space.} \\
\tablenotemark[5]{sum over the first 500 $J=2^{+}~T=0$ states in
the $\Delta N = 2$ space.} \\
\tablenotemark[6]{sum over the first 500 $J=2^{+}~T=1$ states in
the $\Delta N = 2$ space.} \\
\tablenotemark[7]{The experimental value to the $2_{1}^{+}$ state at
4.44 MeV is B(E2)=40 $e^{2}$ fm$^{4}$~\cite{Raman-S:1987ndt}.}
\end{table}

\end{document}